# Diseño de una Arquitectura para la Solución de la Ecuación de Schrödinger usando el Método de Numerov

Victor A. Rodriguez-Toro[§], Fabio Noguera-Leon[‡], Jaime Velasco-Medina[†]

A.A. 25360, Escuela de Ingeniería Eléctrica y Electrónica

Universidad del Valle

Cali, Colombia

[§]victor.rodriguez@correounivalle.edu.co, [‡]fanoleob@univalle.edu.co, [†]jvelasco@univalle.edu.co

*Abstract—* **En este trabajo se presenta una primera aproximación con el propósito de diseñar una arquitectura óptima para la implementación del método de Numerov, el cual permite resolver la Ecuación de Schrödinger Independiente del Tiempo (ESIT) para el caso de una dimensión. El diseño y la simulación han sido realizados mediante el uso de megafunciones de punto flotante de 64 bits disponibles en Quartus II (Versión 9.0) y la verificación de estos resultados fueron realizados usando Matlab. Con los resultados obtenidos, es posible extender este diseño a estructuras paralelas que permitirán calcular varias soluciones de la ESIT.**

## I. Introducción

Los grandes avances que se han desarrollado en los últimos años en la nanotecnología y la nanoelectrónica han planteado importantes desafíos a la comunidad científica en cuanto a la caracterización de nuevos materiales, al diseño de dispositivos a partir de estos materiales y al desarrollo de aplicaciones en general.

Algunos de estos desafíos están relacionados con el hecho de que en la escala nanométrica la materia manifiesta fenómenos que no pueden modelarse mediante las teorías clásicas (e.g. mecánica clásica) y por lo tanto tienen que ser abordados desde la mecánica cuántica. En el marco de esta teoría, las predicciones que pueden realizarse sobre un sistema son distintas a las que se pueden realizar usando las teorías clásicas. Para el caso clásico, un sistema con una posición y una velocidad inicial dada tiene una energía definida para estas condiciones e incluso puede predecirse cuál será su posición final y velocidad final después de un tiempo determinado [1] – [3].

Por otra parte, para el caso cuántico la velocidad no está definida y la evolución de los diferentes sistemas a escala nanométrica sucede en un sentido probabilístico y no determinístico como en las teorías clásicas. Aún más, el sistema nanométrico no está totalmente definido por los parámetros de posición y velocidad, y solo puede encontrarse la probabilidad de que un sistema se encuentre en intervalo espacial dado o la probabilidad de que tenga un rango de valores de momento (análogo cuántico de la velocidad). Los sistemas nanométricos no siguen trayectorias y por la tanto se puede predecir únicamente valores promedios de la posición y los momentos. Adicionalmente, los sistemas a esta escala pueden tunelar, es decir, pueden estar en regiones que clásicamente no sería posible [1-2].

La ecuación que permite hacer estas predicciones es la ecuación de Schrödinger dependiente del tiempo (ESDT) que en general es una ecuación diferencial parcial y no lineal. No obstante, en este trabajo es de especial interés la ecuación de Schrödinger independiente del tiempo (ESIT), la cual es una derivación de la ecuación de Schrödinger más general (ESDT). Resolver la ESIT es análogo a resolver un problema de valores propios donde el conjunto de funciones que satisface la ESIT son las eigenfunciones (estados propios del sistema) y el eigenvalor (o eigenvalores para un caso degenerado [1]) asociado a cada eigenfunción es la energía.

Existen diversos métodos que permiten encontrar la solución de la ESDT y/o la ESIT para una o varias dimensiones tales como el método espectral [3], el método de Numerov [2], entre otros. Sin embargo, para hacer extensivos estos métodos a sistemas más grandes como átomos y moléculas otro tipo de aproximaciones deben ser hechas [2], [4] y los recursos computacionales se hacen cada vez más altos con el incremento del número de coordenadas presentes.

Como una alternativa y primera aproximación para superar la limitación de los recursos computacionales exigidos al solucionar problemas de la mecánica cuántica, en este trabajo se propone una arquitectura en hardware que implementa el método de Numerov para resolver el problema de una partícula sujeto a un potencial parabólico en una dimensión. Este potencial modelo es clave en muchos de los procedimientos de caracterización de nanosistemas (e.g. los modos de vibración de moléculas) [4], o para el estudio de interacción de nanosistemas con el medio. Muchos de estos métodos transforman un problema n-dimensional en n problemas de una dimensión.

El método de Numerov [2], [5], [6] permite calcular las eigenenergías (asociadas a eigenestados ligados únicamente [2]) con una buena aproximación mediante la solución de la ESIT para potenciales unidimensionales arbitrarios discretizados. Esta es una característica ventajosa con respecto a los procedimientos matemáticos donde los potenciales deben aproximarse analíticamente para poder resolver la ESIT. Adicionalmente, este método puede implementarse con relativa facilidad en software pues es una fórmula recursiva (similar a la representación temporal de un filtro digital) cuya operación más costosa es la división. Más aún, existen trabajos a nivel de propuestas de algoritmos en paralelo [7].

Por tratarse de un algoritmo de ensayo y error e inherentemente secuencial, si desean encontrarse varias eigenenergías (un potencial puede tener infinitos posibles valores de eigenenergías), entonces el tiempo de procesamiento será mayor dependiendo el número de energías requeridas.

No se ha encontrado literatura sobre la implementación en hardware para el método de Numerov. No obstante, existen implementaciones para acelerar la simulación de moléculas y lograr establecer sus propiedades estructurales y dinámicas [8]. Muchos de estos trabajos se realizan basados en el paradigma de hardware reconfigurable y con métodos menos exactos que los usados en la mecánica cuántica [9].

Por otra parte, no existe un gran número de trabajos para solucionar la ecuación de Schrödinger por medio de arquitecturas hardware. En este sentido, este trabajo aporta una de las primeras arquitecturas con el objetivo de solucionar la ecuación de Schrödinger para una partícula con potenciales de una dimensión.

Este documento está organizado de la siguiente manera: La Sección II presenta la Ecuación de Schrödinger y el método de Numerov que permite solucionar problemas de la ESIT. La Sección III presenta el hardware implementado así como algunas consideraciones de diseño tenidas en cuenta para este trabajo. En la Sección IV se presentan los resultados obtenidos a partir de la implementación y se comparan en términos de precisión con aquellos calculados por medio de software. La Sección V presenta las conclusiones y los trabajos futuros que pueden ser desarrollados a partir de este trabajo.

## II. MÉTODO DE NUMEROV

### A. Ecuación de Schrödinger

La ecuación de Schrödinger para el caso unidimensional puede describirse como [1-2]

$$i\hbar \frac{\partial \Psi(x,t)}{\partial t} = \hat{H}\Psi(x,t) \quad (1)$$

Donde $i$ es la unidad compleja, $\hbar$ es la constante de Planck dividida $2\pi$, $x$ es la variable espacial, $t$ es la variable temporal y $\Psi$ es la función de onda que contiene toda la información acerca del sistema. $\hat{H}$ es el operador Hamiltoniano el cual contiene toda la información de la energía del sistema y se define como

$$\hat{H} = \hat{T} + \hat{V} \quad (2)$$

Donde $\hat{T}$ es el operador de energía cinética y $\hat{V}$ es el operador de energía potencial. Si el potencial es independiente del tiempo, la ESIT se deriva a partir de (1) y la función $\Psi$ que depende del espacio y el tiempo, se puede escribir como un producto de una función que depende del espacio $\psi$ (eigenfunción) por una función que depende del tiempo $\varphi$. Por medio de la ESIT puede calcularse la función $\psi$ como

$$\hat{H}\psi(x) = E\psi(x) \quad (3)$$

Donde $E$ es la eigenenergía asociada a la eigenfunción. Para el caso específico en que el potencial es parabólico la ecuación de Schrödinger puede escribirse explícitamente como

$$\left[ -\frac{\hbar^2}{2m}\frac{d^2}{dx^2} + kx^2 \right]\psi(x) = E\psi(x) \quad (4)$$

Donde $m$ es un parámetro del sistema y $k$ define la abertura del potencial parabólico.

### B. Método de Numerov

El método de Numerov permite resolver las ecuaciones que tienen la forma como

$$\frac{d^2}{dx^2}\psi(x) = [U(x) + W(x)]\psi(x) \quad (5)$$

Varios modelos matemáticos de sistemas en física permiten aplicar el método de Numerov. Tales modelos son la ESIT, la ecuación de un oscilador armónico clásico sin amortiguamiento y la ecuación de Poisson [5]. Nótese que es posible adoptar la forma de (5) para (4).

El método de Numerov se basa en una descomposición de Taylor de la función a calcular despreciando los órdenes mayores a 5 y relacionando esta expansión con la Ec. (4) [2]. Con estas consideraciones se obtiene una expresión para la función de onda como

$$\psi_{n+1} \approx \frac{\left(2 + \frac{5}{6}g_n s^2\right)\psi_n - \left(1 - \frac{1}{12}s^2 g_{n-1}\right)\psi_{n-1}}{\left(1 - \frac{1}{12}s^2 g_{n+1}\right)} \quad (6)$$

$$g_n = \frac{2m}{\hbar^2}[V_n - E] \quad (7)$$

Donde $s$ es el paso espacial, $n$ es el índice de la iteración, $V_n$ es la función de potencial discretizada y E es la energía deseada del sistema. Para el caso de un potencial armónico y con el objetivo de facilitar el cálculo se multiplica (4) por un factor que está en función de los parámetros $\hbar$, $k$, $m$ [2] lo cual permite tener una versión adimensional de (6) y (7) y que pueden ser escritas como

$$\psi_{n+1} \approx \frac{\left(2 + \frac{5}{6} g_n s^2\right)\psi_n - \left(1 - \frac{1}{12} s^2 g_{n-1}\right)\psi_{n-1}}{1 - s^2 g_{n+1}} \quad (8)$$

$$g_n = 2[V_n - E] = 2[x_n^2 - E] \quad (9)$$

Este algoritmo funciona basado en un principio de prueba y error como se ha mencionado. Debido a que todas las eigenfunciones deben ser bien comportadas, (i.e. convergen a cero en los extremos de la grilla espacial) solo serán válidas aquellas energías $E$ que conserven esta propiedad.

### III. Diseño de Arquitectura Hardware

Para el diseño implementado se ha usado Quartus II y las megafunciones de Altera.

#### A. Ecuación Implementada

Con el objetivo de simplificar los módulos de Hardware requeridos se reescriben (8) y (9) para un potencial cuadrático como

$$\psi_n \approx \frac{1}{12 - s^2 g_n}\left\{\left(24 + 10 s^2 g_{n-1}\right)\psi_{n-1} - \left(12 + - s^2 g_{n-2}\right)\psi_{n-2}\right\} \quad (10)$$

$$g_n = 2[x_n^2 - E] \quad (11)$$

Nótese que (8) ha sido multiplicado y dividido por 12 con el objetivo de reducir el número de divisores requeridos. La forma mostrada en (10) puede ser reescrita de muchas maneras siendo esta una primera aproximación hacia una forma más óptima y en concordancia con el formato numérico de la representación de datos que se usa en este trabajo (IEEE-754 de 64 bits) y el tipo de megafunciones disponibles en ALTERA. Se elige IEEE 754 de 64 bits con el objetivo de disminuir el impacto del error debido al truncamiento, las no linealidades y a la recursividad de (10).

Por otra parte la recursividad depende de $x_n$ como

$$x_{n+1} = x_n + s \quad (12)$$

Y las condiciones iniciales se asumen tales que se conserven las propiedades de buen comportamiento de las eigenfunciones.

$$\psi_{-2} = 0 \quad (13)$$

$$\psi_{-1} = 0.0001 \quad (14)$$

Se ha elegido como extremos de la grilla los valores -5 y 5. No obstante, este valor debe aumentar si se quieren altos valores eigenenergías. El rango elegido en este caso es suficiente para observar hasta diez estados aproximadamente.

$$-5 \leq x \leq 5 \quad (15)$$

#### B. Análisis del Sistema Recursivo

Es posible hacer un análisis del sistema recursivo en (10) y (11) análogo a como se hace con un filtro digital en cuanto a las propiedades de estabilidad, causalidad, linealidad y dinámica (sistema con memoria o sin memoria).

La ecuación (8) muestra que $g_n$ es la entrada y $\psi_n$ la salida. Entonces, como el sistema depende de las muestras pasadas y presentes es dinámico o con memoria. Adicionalmente, el sistema es causal porque no depende de muestras futuras de la entrada o la salida y es no lineal pues la entrada está multiplicada por la salida.

El sistema es inestable en cuanto a que hay funciones que tienden a valores muy grandes por el lado derecho de la grilla. No obstante, debe decirse que esté análisis de estabilidad es más flexible que el que se hace con los filtros digitales.

#### C. Arquitectura Hardware Implementada

Para la implementación de (10) y (11) se requiere de doble precisión pues la recurrencia para el cálculo de la función hace que se propaguen errores numéricos como se ha mencionado. Por lo tanto, se ha usado el formato IEEE-754 de 64 bits para la representación de los datos y bloques de punto flotante de Altera. En la Fig.1 y Fig. 2 se presenta una versión simplificada de la arquitectura digital implementada.

En la Tabla I se presenta un resumen de las megafunciones utilizadas usando el dispositivo EP3C120F780I7 de la familia Cyclone III.

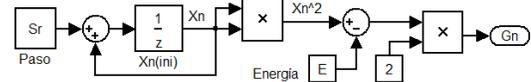

Figura 1. Arquitectura digital implementada para la generación de la entrada.

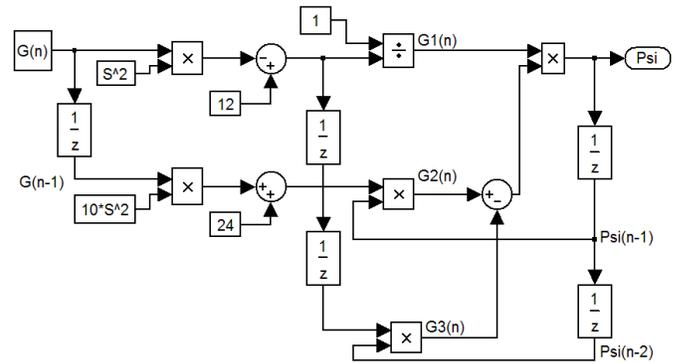

Figura 2. Arquitectura digital implementada para el método de Numerov.

TABLA I
Resumen De los Bloques de Hardware (Megafunciones) Utilizados

| Tipo | Número de Bloques | No. de Ciclos |
|---|---|---|
| Suma | 2 | 7 |
| Resta | 3 | 7 |
| Multiplicación | 6 | 5 |
| División | 1 | 10 |
| MUX | 3 | - |
| Registros | 5 | 1 |
| RAM | 1 | - |

## IV. RESULTADOS, VERIFICACIÓN Y DESEMPEÑO

### A. Resultados de Simulación y Verificación

Como un caso de estudio se calcula la energía del estado fundamental. Para este caso se asume un E=0.5. Se presenta en la Fig. 3 la solución obtenida por el hardware implementado y en la Fig. 4 el valor absoluto de la diferencia de los resultados obtenidos.

Nótese la buena aproximación de los resultados obtenidos por el hardware con relación a los resultados obtenidos usando Matlab. Los errores son del orden de 10E-8 salvo para 3 puntos donde el error es de aproximadamente 10E-3.

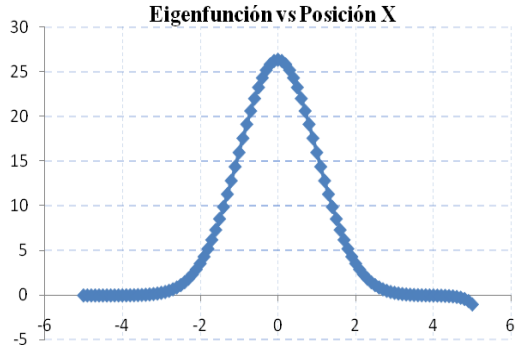

Figura 3. Estado fundamental obtenido en Quartus II.

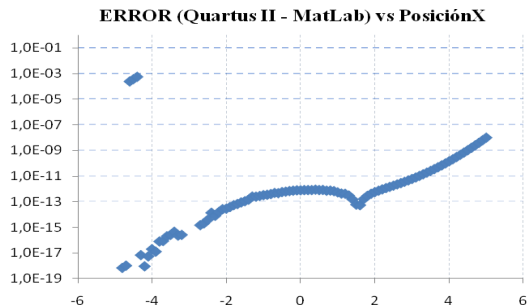

Figura 4. Error de simulación entre Quartus II y MatLab.

### B. Índices de Desempeño

En la Tabla II se presentan los recursos utilizados en la FPGA para la implementación en hardware del método de Numerov.

TABLA II
RECURSOS UTILIZADOS DE LA FPGA

| Recurso | Relación Uso/Disponibilidad | Porcentaje de uso |
|---|---|---|
| ALUTs Combinacionales | 25.773/119.088 | 21.64 |
| Memory bits | 21.400/3.981.312 | 0.54 |
| Bloques DSP de 18 bits | 44/576 | 7.64 |

La frecuencia de operación a la cual se calcularon 100 puntos fue de 20 MHz (i.e. un periodo de 50ns). Primero, se generan los $G_n$ donde cada uno toma 7 ciclos de reloj para generarse (fig. 1). Por lo tanto los 100 puntos de la grilla tardan 35 us.

Después los $G_n$ se introducen al hardware de Numerov (fig. 2), y cada muestra de la salida se ejecuta en 17 ciclos (350 ns). Por lo tanto el cálculo de 98 puntos tarda 83.3 us. Esto implica que todo el cálculo para una sola energía tarda 118,3 us lo cual implica que pueden realizarse 8453 ensayos de posibles energías en un segundo.

Según lo observado en la Tabla II es posible lograr una implementación paralela de más bloques como el mostrado en la Fig. 1 lo cual permitiría ampliar el potencial de este diseño logrando así realizar más cálculos para un tiempo dado.

## V. CONCLUSIONES

La arquitectura hardware presentada en este trabajo cumple con un desempeño satisfactorio en términos de precisión, tiempo de cálculo y recursos de hardware usados. El máximo error obtenido con relación a los datos obtenidos con Matlab fue menor al 10E-3. Por otra parte, el tiempo de cálculo de cada iteración es de 350 ns lo cual puede mejorar sustancialmente con una optimización de los bloques utilizados para esta aplicación. Adicionalmente, los recursos de hardware fueron relativamente bajos acuerdo a los bloques utilizados y a la capacidad de la FPGA utilizada (un máximo 21,64% para todos los recursos).

Una implementación paralela del diseño propuesto permitiría una mayor eficiencia en dos sentidos. En el primero permitiría abarcar una mayor cantidad de valores y en el segundo sentido, para un rango de valores fijos podría hacerse la búsqueda de las energías con una mayor resolución numérica con el objetivo de lograr una mayor precisión.